\documentclass[superscriptaddress,twocolumn,prl]{revtex4}

\usepackage{graphicx}
\usepackage{amsmath}
\usepackage{longtable}
\usepackage{amssymb}
\usepackage{latexsym}
\usepackage{amsmath}
\usepackage{wrapfig}
\usepackage{natbib}

\begin{document}

\title{High-order noise filtering in nontrivial quantum logic gates}

\author{Todd Green}
\affiliation{Centre for Engineered Quantum Systems, School of Physics, The University of Sydney, NSW 2006 Australia}
\author{Hermann Uys}
\affiliation{National Laser Centre, Council for Scientific and Industrial Research, Pretoria, South Africa}
\author{Michael J. Biercuk}
\email[To whom correspondence should be addressed: ]{michael.biercuk@sydney.edu.au}

\affiliation{Centre for Engineered Quantum Systems, School of Physics, The University of Sydney, NSW 2006 Australia}

\date{\today}

\begin{abstract}
Treating the effects of a time-dependent classical dephasing environment during quantum logic operations poses a theoretical challenge, as the application of non-commuting control operations gives rise to both dephasing and depolarization errors that must be accounted for in order to understand total average error rates.  We develop a treatment based on effective Hamiltonian theory that allows us to efficiently model the effect of classical noise on nontrivial single-bit quantum logic operations composed of arbitrary control sequences.  We present a general method to calculate the ensemble-averaged entanglement fidelity to arbitrary order in terms of noise filter functions, and provide explicit expressions to fourth order in the noise strength.  In the weak noise limit we derive explicit filter functions for a broad class of piecewise-constant control sequences, and use them to study the performance of dynamically corrected gates, yielding good agreement with brute-force numerics. 
\end{abstract}

\pacs{}

\maketitle

Dynamical error suppression strategies have been demonstrated as a means by which errors due to decoherence may be suppressed during qubit memory operations~\cite{Viola1998,Viola1999, Zanardi1999,Vitali1999, Byrd2003,Khodjasteh2005, Yao2007, Kuopanportti2008}.  In such cases, a filter-design framework~\cite{Martinis2003, Uhrig2007, Cywinski2008, Biercuk2009, Biercuk_Filter} has successfully shown how to precisely estimate average error rates in the presence of classical, time-dependent noise.  Expanding this analysis beyond the identity operator has proved challenging due to the need for efficient techniques to treat a random, time-varying noise term that does not commute with the applied control field.  Understanding the influence of such time-dependent processes during control operations is vital, however, as environmental decoherence sets the lower-bound on achievable gate error rates in a quantum informatic setting.

In this letter we address the challenge of characterizing and mitigating decoherence due to classical noise during nontrivial single-qubit operations.  We calculate the ensemble-averaged entanglement fidelity for an arbitrary control sequence to fourth order in the noise strength, incorporating terms to the third order of the Magnus Expansion in an effective Hamiltonian treatment.   For concreteness we explicitly calculate the filter functions to lowest nontrivial order for sequences composed of $\pi$-rotations about Cartesian axes with arbitrary rotation rates - a class including dynamically corrected gates (DCGs).  Our results permit detailed calculation of the generic (amplitude and phase) errors that result from applying a quantum control operation in the presence of pure-dephasing noise, and validate perturbative predictions of an increased order of error suppression~\cite{khodjasteh2009dcg, khodjasteh2009pradcg}.  This straightforward analytical approach is compared against brute-force numerical calculations of the evolution of the Bloch vector, and shows excellent agreement.

We consider the canonical dephasing environment in which the system evolves freely under a Hamiltonian of the form $H=\frac{1}{2}[\Omega+\eta(t)]\sigma_{z}$, where $\Omega$ is the unperturbed qubit splitting, $\eta$ is a time-dependent classical random variable, and $\sigma_{z}$ is a Pauli operator.  In the case of free-evolution the presence of a nonzero $\eta(t)$ produces dephasing in an ensemble average.  However, during driven operations where one applies a control field proportional to $\sigma_{x}$ or $\sigma_{y}$, the presence of a pure-dephasing noise environment yields both polarization damping and dephasing effects.  Both must be considered in a full treatment of gate-errors.  This accounts for the most significant \emph{correctable} forms of decoherence in experiment; most remaining polarization-damping errors are due to stochastic processes that cannot be corrected through dynamical error suppression.

We begin with an outline of our method. The total Hamiltonian (in the rotating frame at $\Omega$) is
$H(t)=H_{0}(t)+H_{c}(t)$, over $t\in[0,\tau]$.  The operator $H_{0}(t)=\eta(t)\sigma_{z}/2$ represents a time-varying dephasing environment, while $H_{c}(t)$ describes an interaction between the system and an external control device that, in principle, may be used to implement arbitrary rotations of the Bloch vector. In general, evaluating the total propagator $U(t)=T\exp(-i\int_{0}^{t}H(t')dt')$ explicitly for $H_{0}(t)\neq0$ is difficult due to noncommuting terms in $H(t)$. We therefore proceed by factoring out that part of the qubit evolution that is due solely to the control and expressing the residual `error propagator' $\tilde{U}(\tau)$ in terms of a \emph{time-independent} effective Hamiltonian that can then be evaluated, following a general procedure laid out in Ref.~\cite{Viola1999}.

Defining the propagator $U_{c}(t)=T\exp(-i\int_{0}^{t}H_{c}(t')dt')$, it can be shown that $\tilde{U}(t)= U_{c}^{\dag}(t)U(t)$ satisfies the equation of motion $i\frac{d\tilde{U}(t)}{dt}=\tilde{H}_{0}(t)\tilde{U}(t)$, where $\tilde{H}_{0}(t)\equiv U_{c}^{\dag}(t)H_{0}(t)U_{c}(t)$ ~\cite{HaeberlenWaugh, Haeberlen1976}. If $H_{c}(t)$ enacts a target unitary operation $Q$ in the absence of decoherence, then we may write the total noise-affected operation as $U(\tau)=Q \tilde{U}(\tau)$.
The error propagator can then be expressed in terms of a time-independent effective Hamiltonian $\overline{H}$, defined by $\tilde{U}(\tau)\equiv\exp(-i\overline{H}\tau)$.

For a dephasing Hamiltonian we have $\tilde{H}_{0}(t)=\frac{\eta(t)}{2}U_{c}^{\dag}(t)\sigma_{z}U_{c}(t)$; the term $U_{c}^{\dag}(t)\sigma_{z}U_{c}(t)$ is a rotation of the $\sigma_{z}$ operator due to the control.  We may therefore define a time-dependent `control vector' in a Cartesian basis ($l=x,y,z$), $\vec{s}_{1}(t)=\sum_{l}s_{1,l}(t) \hat{l}$, where $|\vec{s}_{1}(t)|=1$,  $\forall t\in[0,\tau]$, such that $\tilde{H}_{0}(t)=\frac{\eta(t)}{2}\vec{s}_{1}(t)\cdot\vec{\sigma}$. Since $\tilde{H}_{0}(t)$ belongs to the Lie algebra $su(2)$, $\forall t\in[0,\tau]$,  the effective Hamiltonian $\overline{H}$ derived from it is also in $su(2)$. 

We may now write $U(\tau)=Q\exp[-i\vec{a}(\tau)\cdot\vec{\sigma}]$, where the effect of the noise on the ideal operation $Q$ is encoded in the `error vector' $\vec{a}(\tau)$. Here  $\vec{a}(\tau)=\sum_{l}a_{l}(\tau)\hat{l}$ represents error contributions to $Q$ due to terms in $\overline{H}$ proportional to the Cartesian components of the Pauli operator.  In general, the action of the control converts pure-dephasing noise to arbitrary rotations of the qubit's Bloch vector, producing both dephasing and polarization damping errors.

In most cases, an explicit expression for $\exp[-i\vec{a}(\tau)\cdot\vec{\sigma}]$ can only be found using approximation methods. For our purposes, it is desirable that the simple exponential form of the operator be retained to any level of approximation. This can be achieved through the use of the Magnus expansion~\cite{MagnusExpansion, MagnusExpansion2}. Using this method, the exponent is expanded in an infinite series of time-integrals over nested commutators of $\tilde{H}_{0}(t)$ at different times. In conjunction with the identity $[\vec{u}\cdot\vec{\sigma},\vec{v}\cdot\vec{\sigma}]=2i(\vec{u}\times\vec{v})\cdot\vec{\sigma}$, $\vec{u},\vec{v}\in \mathbb{R}^{3}$, we find that we can write $\vec{a}(\tau)\cdot\vec{\sigma}=\sum_{i=1}^{\infty}\vec{a}_{i}\cdot\vec{\sigma}$ to all orders $i$. The first three terms in series expansion of the error vector are $\vec{a}_{1}(\tau)=\frac{1}{2}\int_{0}^{\tau}dt\eta(t)\vec{s}_{1}(t)$, $\vec{a}_{2}(\tau)=\frac{1}{4}\int_{0}^{\tau}dt_{2}\int_{0}^{t_{2}}dt_{1}\eta(t_{1})\eta(t_{2})\vec{s}_{2}(t_{1},t_{2})$, and $\vec{a}_{3}(\tau)=\frac{1}{12}\int_{0}^{\tau}dt_{3}\int_{0}^{t_{3}}dt_{2}\int_{0}^{t_{2}}dt_{1}
\eta(t_{1})\eta(t_{2})\eta(t_{3})\vec{s}_{3}(t_{1},t_{2},t_{3})$. Here, high-order commutators in the Magnus expansion have been reduced to vector cross products of the control vector at different times, $\vec{s}_{2}(t_{1},t_{2})\equiv\vec{s}_{1}(t_{2})\times\vec{s}_{1}(t_{1})$ and $\vec{s}_{3}(t_{1},t_{2},t_{3})\equiv\vec{s}_{1}(t_{3})\times\left[\vec{s}_{1}(t_{2})\times\vec{s}_{1}(t_{1})\right]+\left[\vec{s}_{1}(t_{3})\times\vec{s}_{1}(t_{2})\right]\times\vec{s}_{1}(t_{1})$.

We evaluate the net effect of the gate operation $Q$ in the presence of noise via the ensemble average entanglement fidelity~\cite{EntanglementFidelity}, $\left\langle \mathcal{F}(\tau)\right\rangle=\left\langle\left|\textrm{Tr}\left(QU(\tau)/2\right)\right|^{2}\right\rangle$ which, when written in terms of the error vector, becomes  $\langle\mathcal{F}(\tau)\rangle=\frac{1}{2}\left[\langle\cos[2|\vec{a}|]\rangle+1\right]$. To evaluate the fidelity, we write  $|\vec{a}|=(\sum_{l}a_{l}^2)^{1/2}$, express the cosine term as a Taylor series and substitute the Magnus expansion $a_{l}=\sum_{i=1}^{\infty}a_{i,l}$ for each component of the error vector. The result is an infinite series of multi-dimensional integrals over products of multiple-time noise correlation functions and components of the control vector $\vec{s}_{1}(t)$. For example, the lowest order effects of the noise are captured by $\langle a_{1, l}^2\rangle =\int_{-\infty}^{\infty}\int_{-\infty}^{\infty}s_{1,l}(t_{1})s_{1,l}(t_{2})\left\langle\eta(t_{1})\eta(t_{2})\right\rangle dt_{1}dt_{2}$,  for $l=x,y,z$.

Assuming the noise is Gaussian, only correlation functions $\langle\eta(t_1)...\eta(t_{n})\rangle$ for which $n$ is even contribute. Further, applying the Gaussian moment theorem, each of these can be written in terms of simple \emph{two-point} correlation functions. Using the root mean square deviation of the noise, $\Delta\eta\equiv\sqrt{\langle\eta(t)^{2}\rangle}$, as a measure of the noise strength, we can define a parameter $\xi\equiv\Delta\eta\tau/2$ which provides an upper bound for the magnitude of each term in the series expansion of the cosine function~\cite{Viola:BADD,notexi}. If we restrict our analysis to weak-noise/ efficient-control conditions under which $\xi<1$, then higher-order terms provide diminishing contributions to the total error and we may truncate the series. To fourth order in $\xi$ we find that

\begin{align}
&\langle\cos[2|\vec{a}|]\rangle=1-2\xi^{2}\left\{\langle \tilde{a}_{1,l}^{2}\rangle\right\}\notag\\
&-2\xi^{4}\left\{\frac{1}{4}\left(3\langle \tilde{a}_{2,l}^{2}\rangle+2\langle \tilde{a}_{1,l}\tilde{a}_{3,l}\rangle\right)-\langle \tilde{a}_{1,l}^{2}\tilde{a}_{1,l'}^{2}\rangle\right\}
\label{Eq:CosExpansion}
\end{align}

\noindent where sums are implicitly performed over $l$ and $l'$, $l\neq l'$, and the tilde indicates that the magnitude of the integrals has been factored out ($-1\leq \tilde{a}_{i,l}\leq1$).  This expression includes terms to \emph{third} order in the Magnus expansion as they contribute to the same order in $\xi$ as the second order term.

The various contributions to Eq.~\ref{Eq:CosExpansion} may be calculated explicitly by Fourier transforming the noise and control.  For instance, we may write
\begin{align}
\langle a_{1, l}^2\rangle 
=\frac{1}{4\pi}\int_{0}^{\infty}\left|y_{1,l}(\omega)\right|^{2}\frac{S(\omega)}{\omega^{2}}d\omega.
\end{align}
\noindent with $y_{1,l}(\omega)=-i\omega\int_{-\infty}^{\infty}s_{1,l}(t) e^{i\omega t}dt$, capturing terms proportional to $\sigma_{l}$. We sum over $l$ to write $F_{1}(\omega)=\sum_{l}F_{1,l}=\sum_{l}\left|y_{1,l}(\omega)\right|^{2}$, corresponding to the terms in the first line of Eq.~\ref{Eq:CosExpansion}.  Similarly, we account for the proliferation of higher order terms arising from the vector cross product by defining $F_{p,2}(\omega,\omega', \tau)$, where $p$ is an index over terms proportional to $\xi^{4}$ in Eq.~\ref{Eq:CosExpansion}.  These terms contain four-point correlation functions in time, $\left\langle\eta(t_{1})\eta(t_{2})\eta(t_{3})\eta(t_{4})\right\rangle$, which may be explicitly evaluated using the Gaussian moment theorem.  

We may then compactly write the entanglement fidelity

\begin{widetext}
\begin{align}
\left\langle \mathcal{F} \right\rangle&=1-\frac{1}{4\pi}\int_{0}^{\infty}\frac{d\omega}{\omega^2} S(\omega)F_{1}(\omega,\tau)
-\frac{1}{(4\pi)^{2}}\sum_{p}\int_{0}^{\infty}\frac{d\omega}{\omega^2}S(\omega)\int_{0}^{\infty}\frac{d\omega'}{\omega'^2}S(\omega')F_{p,2}(\omega,\omega',\tau).
\label{eq:highorderfilter}
\end{align}
\end{widetext}
\noindent Terms to all orders in $\xi$ may be evaluated using the same procedure.

With these expressions we have reduced the effect of a time-dependent dephasing environment and a time-dependent control Hamiltonian to integrals incorporating only stationary statistical properties of the system: the noise power spectral density and the spectral functions to arbitrary order, $F_{i}$.  These terms contain all relevant information about the applied control, and the explicit inclusion of terms proportional to all $\sigma_{l}$ captures both dephasing and polarization-damping errors produced during control operations.  We refer to the terms $F_{i}$ as the \emph{filter functions} for the total control operation, in analogy with previous work on dynamical decoupling~\cite{Uhrig2007, Cywinski2008, Biercuk2009, Biercuk_Filter}.  The leading nontrivial term is closely related to the idea of spectral overlap functions between control and noise that has been studied previously~\cite{Kofman2004, Gordon2008, Clausen2010} , but a generalized derivation to high order has not appeared to the best of our knowledge.

We now consider a specific case for concreteness that is germane to many coherent control experiments, including the implementation of dynamically corrected gates.  We define $H_{c}(t)$ as piecewise-constant such that, during the $j$-th interval, the control is intended to execute $\sigma_{l_{j}}$, restricted here to either the identity $I$, or a rotation of the qubit Bloch vector through $\pm\pi$ about one of the three Cartesian directions.  In this notation, $l_{j}=I,x,y,z$.

The control propagator may be written explicitly such that during the $j$th driven operation we have $U_{c}^{(j)}=\exp[-i\Omega_{R}^{(j)}(t-t_{{j-1}})\sigma_{l_{j}}/2]\sigma_{\forall j-1}$, where $\Omega_{R}^{(j)}$ gives the driven rotation rate about axis $\hat{l}$ in time-bin $j$.  The operator $\sigma_{\forall j-1}=\sigma_{l_{j-1}}\sigma_{l_{j-2}}...\sigma_{l_{1}}$ describes the cumulative effect of all completed rotations in the preceding time segments.

We restrict our presentation to terms in the entanglement fidelity of order $\xi^{2}$, appropriate for the case of weak dephasing noise.  Higher order contributions are straightforward to calculate explicitly, but involve many dozens of terms.  In this case we only require first-order components of the error vector and may approximate $\left\langle\mathcal{F(\tau)}\right\rangle\approx \frac{1}{2}\left[e^{-2\sum \langle a_{1,l}^2\rangle}+1\right]$.  Using the first-order filter functions we have $\langle \mathcal {F(\tau)}\rangle\approx\frac{1}{2}\left[\exp(-\chi(\tau))+1\right]$, where $\chi(\tau)=\frac{1}{2\pi}\int_{0}^{\infty}F_{1}(\omega)\frac{S(\omega)}{\omega^{2}}d\omega$.

For piecewise-constant control as described above, the terms contributing to $F_{1,l}(\omega)$ may be written:
\begin{align}
\nonumber y_{1,x}(\omega)&=\sum_{j=1}^{k}(-1)^{N^{[j]}_{y,z}+1}\frac{i\omega\Omega^{(j)}_{R}}{\omega^{2} -\left(\Omega^{(j)}_{R}\right)^{2}} \left(e^{i\omega t_{j}}+e^{i\omega t_{j-1}}\right)\delta_{l_{j}y}\\\notag
\nonumber y_{1,y}(\omega)&=\sum_{j=1}^{k}(-1)^{N^{[j]}_{x,z}}\frac{i\omega\Omega^{(j)}_{R}}{\omega^{2}-\left(\Omega^{(j)}_{R}\right)^{2}}\left(e^{i\omega t_{j}}+e^{i\omega t_{j-1}}\right)\delta_{l_{j}x}\\\notag
\end{align}

\begin {widetext}
\begin{align}
y_{1,z}(\omega)&=\sum_{j=1}^{k}(-1)^{N^{[j]}_{x,y}}\left[\frac{\omega^{2}}{\omega^{2}-\left(\Omega^{(j)}_{R}\right)^{2}}
\left(e^{i\omega t_{j}}+e^{i\omega t_{j-1}}\right)(\delta_{l_{j}x}+\delta_{l_{j}y})+\left(e^{i\omega t_{j-1}}-e^{i\omega t_{j}}\right)(\delta_{l_{j}z}+\delta_{l_{j}I})\right]\label{Eq:Filters}
\end{align}
\end{widetext}

\noindent where $\delta_{\alpha\beta}=1$ for $\alpha=\beta$ and is zero otherwise.  Here $N_{\alpha,\beta}^{[j]}$ represents the number of times $\sigma_{\alpha}$ or $\sigma_{\beta}$ appear in the sequence up to the $j$-th interval.  We note that the prefactors in these equations are similar to those derived from the steady-state master-equation treatment of a driven two-level system in the presence of dissipation~\cite{OpenQuantumSystems}.

\begin{figure}[bp]
\includegraphics[width=6.5cm]{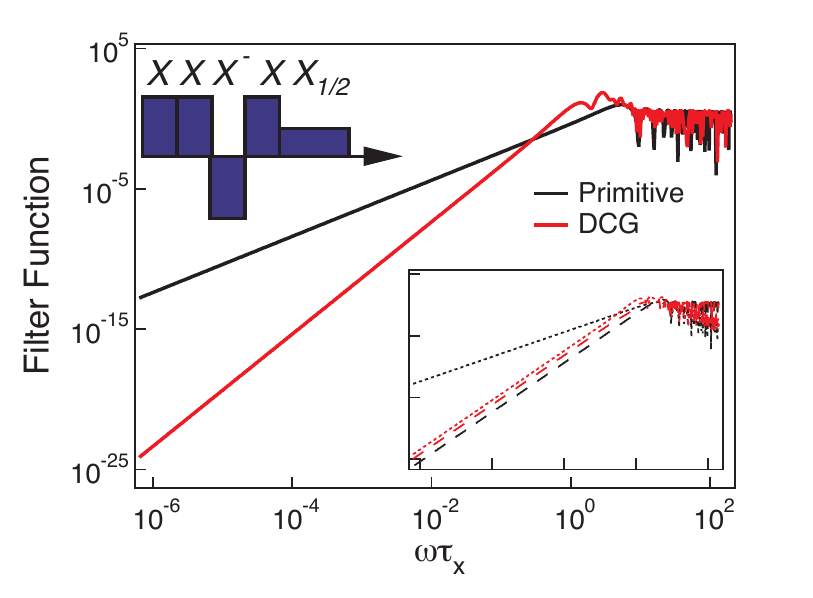}\\
\caption{\label{Fig:F2} (Color online) Filter functions for primitive $X$ and $X_{DCG}$, based on Eq.~(\ref{Eq:Filters}).  Upper inset, schematic construction of $X_{DCG}$~\cite{khodjasteh2009dcg}.  $X^{-}$ is a $X$ rotation with $\pi$ phase shift.  Lower inset, Amplitude, $F_{1,x}$ (dotted), and Phase $F_{1,z}$ (dashed) filter functions for the same gates, denoted by color.  Tick marks same as main panel. }
\end{figure}

Using this approach we are able to analyze the effects of classical noise on a complete set of both primitive (standard) and dynamically protected single-qubit gates of interest for quantum logic.  As an example, in Fig.~\ref{Fig:F2} we show the total filter functions derived from this treatment for $X$ operations in primitive and DCG formulations~\cite{Viola2003, khodjasteh2009dcg}.  Consistent with previous perturbative treatments of quantum-mechanical baths, we find that the order of error suppression, given by the low-frequency rolloff of $F_{1}(\omega)$, is increased in the DCG relative to the primitive gate~\cite{Biercuk_Filter, Hayes_WDD}.  The extension of the duration of $X_{DCG}$ relative to $X$ is manifested as a decrease in $\omega_{F1}$, the frequency above which the filter function takes value unity and noise is passed largely unimpeded.   By examining the phase and amplitude components of the filter function independently ($F_{1,z}$ and $F_{1,x}$), we see that for a DCG we improve the order of error suppression primarily in the term that commutes with the control operation (e.g. amplitude for an $X$ gate).

The form of the filter functions for other operations and their DCG constructions (e.g. $Z_{\theta}$, $H$) show similar behavior. Dynamical decoupling (dynamically protected $I$) may also be treated using Eq.~(\ref{Eq:Filters}), and will be addressed in detail in a separate manuscript. 

\begin{figure}[tp]
\includegraphics[width=8cm]{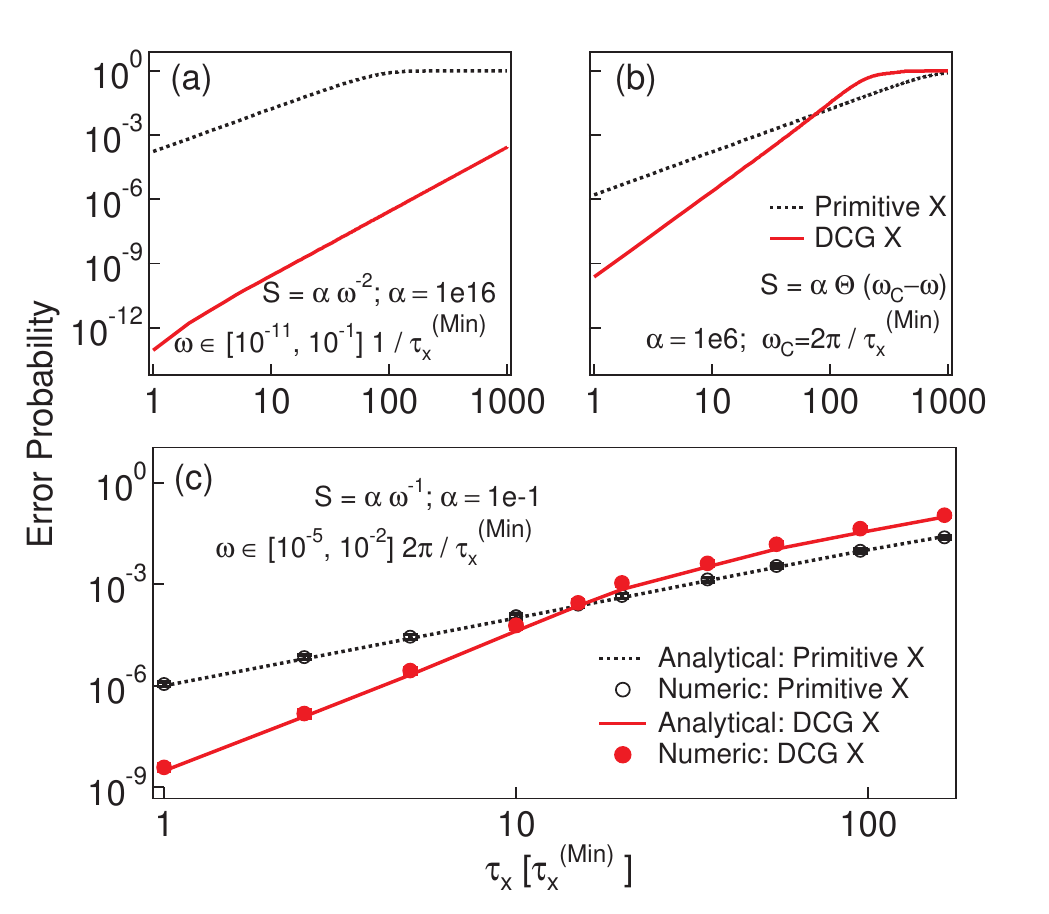}\\
\caption{\label{Fig:F3} (Color online) Calculated error rates for primitive $X$ and $X_{DCG}$ gates in the presence of noise.  (a) Calculate error rates in the presence of noise similar to that observed in Ref.~\cite{BiercukPRB2011}, appropriate for decoherence due to nuclear spin diffusion in solid-state singlet-triplet qubits.  Data are plotted in dimensionless units of the minimum $\tau_{x}$ value.  (b) Similar to (a), but using a white noise power spectral density up to a sharp high-frequency cutoff.  (c) Comparison of calculated error rates based on analytical filter functions derived herein and brute-force numerics.  Noise strength in numerics is set using $\Delta=0.008\; \left(\tau^{(\text{Min})}_{x}\right)^{-1}$, guaranteeing convergence of the Magnus expansion for this range of $\tau_{x}$.  Each data point averages over 50 randomly generated noise trajectories whose statistical properties reproduce $S(\omega)$.  Error bars are derived from the root-mean-square deviation of the individual trajectory results, but are small compared to the marker size.}
\end{figure}

Using the lowest-order approximation for the entanglement fidelity and the specific forms of the filter functions for  $X_{DCG}$ we calculate the error probability for different noise environments and plot these in Fig.~\ref{Fig:F3}.  As expected, in an environment given by $S(\omega)=\alpha/\omega^{2}$, with $\alpha$ a scaling factor (c.f. Ref~\cite{BiercukPRB2011}), we find significant benefits from using the DCG construction.  By contrast, in a white noise environment with a sharp high-frequency cutoff ($S(\omega)=\alpha\Theta(\omega_{c}-\omega)$) the significant high-frequency spectral components of the noise and the extended duration of the DCG construction can yield a net performance degradation in the event of long $\tau_{x}$ and large $\omega_{c}$. The relationship between these two quantities thus provides a dominant practical limit on the applicability of DCG construction in realistic settings; so long as $\omega_{c}\lesssim\tau_{x}^{-1}$ the DCG provides net performance enhancement.

Validation for the assumptions and approximations made in this treatment comes from performing detailed brute-force numerical simulation of the evolution of the Bloch vector in the presence of a noisy environment, and averaging over multiple trajectories, $\eta(t)$ (Fig.~\ref{Fig:F3}c).  The trajectories are chosen to exhibit the statistical properties of a desired power spectral density.  Calculations involving only the terms in Eq.~\ref{Eq:Filters} accurately reproduce numerically calculated error probabilities to within $\sim20\%$ for the $X$ gate.  In complex DCG sequences efficient decoupling reduces leading-order terms until residual contributions become comparable to higher-order terms.  We observe that the lowest-order filter functions underestimate error and deviation from numerics grows with more complex sequences or stronger noise, but remains a factor of order unity for the cases we have studied.

In summary we have developed a theoretical treatment permitting the calculation of error rates due to time-varying classical noise during arbitrary control operations, to arbitrary order.  We have explicitly produced a general high-order approximation to the qubit's entanglement fidelity and using the first three orders of the Magnus expansion.  Building on these results we have presented an example case of piecewise-constant control, and given simple, leading-order expressions for the filter function.  This allows any experimentalist to evaluate expected error rates to the correct order of magnitude \emph{during nontrivial control operations}, accounting for the fact that pure dephasing noise can result in both dephasing and polarization damping errors during control operations.  Included in this class of control sequences are dynamically corrected gates, and we have validated previous perturbative calculations for DCG performance through an intuitive noise-filtering approach.

 We believe that this experimentally accessible and rigorously tested method for understanding the influence of classical noise in single-qubit logic operations will prove valuable to the community and will open new pathways for the development of error-robust quantum control strategies.   We emphasize that our approach is technology independent and these methods apply to any quantum system.

\begin{acknowledgments}
\textit{Acknowledgements:}  The authors thank A.C Doherty, K. Khodjasteh, and L. Viola for useful discussions.  This work partially supported by the US Army Research Office under Contract Number  W911NF-11-1-0068, and the Australian Research Council Centre of Excellence for Engineered Quantum Systems CE110001013.
\end{acknowledgments}


\begin{thebibliography}{29}
\bibitem[{not()}]{notexi}
\bibinfo{note}{If the power spectral density $S(\omega)$ of the noise has a
  sharp high-frequency cutoff $\omega_{c}$, then $\Delta\eta$ becomes an
  increasing function of the cutoff frequency,
  $\Delta\eta=\Delta\eta(\omega_{c})$. Additionally, the total sequence time
  $\tau$ is inversely related to the minimum value of the bandwidth
  $\Omega_{\text{ctrl}}$ of the control ~\cite{Viola:BADD}. The parameter $\xi$
  can then be written as $\xi=c\Delta\eta(\omega_{c})/\Omega_{\text{ctrl}}$,
  where $c$ is a constant. For example, in the case of of white noise,
  $S(\omega)=\alpha\Theta(\omega_{c}-\omega)$, we have
  $\xi=c\alpha\sqrt{\omega_{c}}/\Omega_{\text{ctrl}}$. Thus the requirement
  that $\xi\ll1$ corresponds to a the regime in which the control is
  sufficiently `fast' relative to the noise.}
\end{thebibliography}

\end{document}


\maketitle
\nocite{*}
\renewcommand{\baselinestretch}{1.3}

\section{Introduction}
In the main body of the manuscript we showed, through Equation (1), how one may write a general expression for the entanglement fidelity under dephasing noise and arbitrary control to fourth order in a parameter $\xi$
\begin{equation}\label{fidapp}
\langle\mathcal{F}\rangle\simeq1-\sum_{i}\langle a_{1,i}^2\rangle
-\left\{\sum_{i}\left(\langle a_{2,i}^2\rangle+2\langle a_{1,i}a_{3,i}\rangle\right)
-\frac{1}{3}\sum_{i,j}\langle a_{1,i}^{2}a_{1,j}^{2}\rangle\right\}
\end{equation}

We remind the reader that the error vector can then be written as an infinite sum $\vec{a}=\vec{a}_{1}+\vec{a}_{2}+...$, with numbered subscripts on the $a$ terms refer to the order of the Magnus expansion from which they are derived.  The first three terms take the form:
\begin{align}\label{magterms}
\begin{split}
\vec{a}_{1}(\tau)=&\frac{1}{2}\int_{0}^{\tau}dt\eta(t)
\vec{s}_{1}(t)\\
\\
\vec{a}_{2}(\tau)=&\frac{1}{4}\int_{0}^{\tau}dt_{2}\int_{0}^{t_{2}}dt_{1}
\eta(t_{1})\eta(t_{2})\vec{s}_{2}(t_{1},t_{2})\\
\\
\vec{a}_{3}(\tau)=&\frac{1}{12}\int_{0}^{\tau}dt_{3}\int_{0}^{t_{3}}dt_{2}\int_{0}^{t_{2}}dt_{1}
\eta(t_{1})\eta(t_{2})\eta(t_{3})\vec{s}_{3}(t_{1},t_{2},t_{3})\\
\end{split}
\end{align}
where
\begin{align}
\vec{s}_{2}(t_{1},t_{2})&\equiv\vec{s}_{1}(t_{2})\times\vec{s}_{1}(t_{1})\\\nonumber
\vec{s}_{3}(t_{1},t_{2},t_{3})&\equiv\vec{s}_{1}(t_{3})\times\left(\vec{s}_{1}(t_{2})\times\vec{s}_{1}(t_{1})\right)
+\left(\vec{s}_{1}(t_{3})\times\vec{s}_{1}(t_{2})\right)\times\vec{s}_{1}(t_{1}).
\end{align}

The fidelity may then ultimately be expressed in terms of noise power spectral density as 
\begin{align}
\left\langle \mathcal{F} \right\rangle&=1-\frac{1}{4\pi}\int_{0}^{\infty}\frac{d\omega}{\omega^2} S(\omega)F_{1}(\omega,\tau)
-\frac{1}{(4\pi)^{2}}\sum_{p}\int_{0}^{\infty}\frac{d\omega}{\omega^2}S(\omega)\int_{0}^{\infty}\frac{d\omega'}{\omega'^2}S(\omega')F_{p,2}(\omega,\omega',\tau).
\label{eq:highorderfilter}
\end{align}

In this supplementary material we describe in detail how to calculate the various terms $F_{p,2}(\omega,\omega',\tau)$.  As the index $p$ generically references higher order contributions to the filter function we will not use this index explicitly in what follows. In all cases notation follows conventions and definitions introduced in the main text.

\section{Details of Terms in Equation (\ref{fidapp}):}

\subsection{$\langle a_{1,i}a_{1,j}\rangle$ - Includes $\langle a_{1,i}^{2}\rangle$ as a Special Case ($i=j$)}
(\emph{Though only $\langle a_{1,i}^{2}\rangle$ appears in equation (\ref{fidapp}), the more general result will be required for other terms})
\begin{equation}\label{11}
\langle a_{1,i}a_{1,j}\rangle=\frac{1}{4}\int_{0}^{\tau}dt_{2}s_{1,i}(t_{2})\int_{0}^{\tau}dt_{1}s_{1,j}(t_{1})
\langle\eta(t_{1})\eta(t_{2})\rangle
\end{equation}

\subsubsection{Bound}
Using the rms value of the noise and the definition for 
$\Delta\eta=\sqrt{\langle\eta(t)^2\rangle}$, we know $\left|s_{1,i}(t_{2})s_{1,j}(t_{1})\right|\leq 1$, and therefore
\begin{equation}
\left|\langle a_{1,i}a_{1,j}\rangle\right|\leq\frac{\Delta\eta^2\tau^2}{4}\equiv\xi^2
\end{equation}

\subsubsection{Explicit form in terms of the noise power spectrum}

Returning to equation (\ref{11}), we write $g(t_2-t_1)\equiv\langle\eta(t_1)\eta(t_2)\rangle$
in terms of the power spectral density $S(\omega)$, ie.
\begin{equation}
g(t_2-t_1)=\frac{1}{2\pi}\int_{-\infty}^{\infty}S(\omega)e^{i\omega (t_2-t_1)}
\end{equation}
so that
\begin{align}
\langle a_{1,i}a_{1,j}\rangle&=\frac{1}{8\pi}\int_{-\infty}^{\infty}\frac{d\omega}{\omega^2} S(\omega)
y_{1,i}(\omega,\tau)y_{1,j}(\omega,\tau)^{*}\\\nonumber
&=\frac{1}{4\pi}\int_{0}^{\infty}\frac{d\omega}{\omega^2} S(\omega)
Re\left[y_{1,i}(\omega,\tau)y_{1,j}(\omega,\tau)^{*}\right]
\end{align}
where
\begin{equation}
y_{1,i}(\omega,\tau)=-i\omega\int_{0}^{\tau}dt s_{1,i}e^{i\omega t}
\end{equation}
\noindent Note the implicit difference between the subscript index $i$, and the imaginary number.

For $i=j$
\begin{align}
\langle a_{1,j}^{2}\rangle=\frac{1}{4\pi}\int_{0}^{\infty}\frac{d\omega}{\omega^2} S(\omega)
\left|y_{1,j}(\omega,\tau)\right|^{2}
\end{align}
Summing over all cartesian coordinates in Eq.~\ref{fidapp} above, this allows us to write the contribution dependent on $F_{1}(\omega)$.

\newpage
\subsection{$\langle a_{1,i}^{2}a_{1,j}^{2}\rangle$ - Includes $\langle a_{1,j}^{4}\rangle$ as a Special Case ($i=j$)}

\begin{equation}\label{21}
\langle a_{1,i}^{2}a_{1,j}^{2}\rangle=\frac{1}{16}\int_{0}^{\tau}dt_{4}s_{1,i}(t_{4})\int_{0}^{\tau}dt_{3}s_{1,i}(t_{3})
\int_{0}^{\tau}dt_{2}s_{1,j}(t_{2})\int_{0}^{\tau}dt_{1}s_{1,j}(t_{1})
\langle\eta(t_{1})\eta(t_{2})\eta(t_{3})\eta(t_{4})\rangle
\end{equation}

\subsubsection{Bound}
Following the same general arguments as above we find an upper bound
\begin{equation}
\langle a_{1,i}^{2}a_{1,j}^{2}\rangle\leq\frac{3\Delta\eta^4\tau^4}{16}=3\xi^4
\end{equation}

\subsubsection{Explicit form in terms of the noise power spectrum}
We begin by applying the Gaussian moment theorem
\begin{align}\label{GMT}
\langle\eta(t_{1})\eta(t_{2})\eta(t_{3})\eta(t_{4})\rangle=&
\langle\eta(t_{1})\eta(t_{2})\rangle\langle\eta(t_{3})\eta(t_{4})\rangle\notag\\
+&\langle\eta(t_{1})\eta(t_{3})\rangle\langle\eta(t_{2})\eta(t_{4})\rangle
+\langle\eta(t_{1})\eta(t_{4})\rangle\langle\eta(t_{3})\eta(t_{2})\rangle
\end{align}
we have
\begin{equation}
\langle a_{1,i}^{2}a_{1,j}^{2}\rangle=\langle a_{1,i}^2\rangle\langle a_{1,j}^2\rangle+2\langle a_{1,i}a_{1,j}\rangle^2.
\end{equation}

We may therefore write this term as
\begin{equation}
\langle a_{1,i}^{2}a_{1,j}^{2}\rangle=\frac{1}{(4\pi)^2}
\int_{0}^{\infty}\frac{d\omega}{\omega^2}S(\omega)\int_{0}^{\infty}\frac{d\omega'}{\omega'^2}S(\omega')
F_{2,c,i,j}(\omega,\omega',\tau)
\end{equation}
where
\begin{align}
F_{2,c,i,j}(\omega,\omega',\tau)
\equiv\left|y_{1,i}(\omega,\tau)\right|^{2}\left|y_{1,j}(\omega',\tau)\right|^{2}
+2Re\left[y_{1,i}(\omega,\tau)y_{1,j}(\omega,\tau)^{*}\right]
Re\left[y_{1,i}(\omega',\tau)y_{1,j}(\omega',\tau)^{*}\right]
\end{align}

For $i=j$ we use $\langle a_{1,j}^{4}\rangle=3\langle a_{1,j}^{2}\rangle^2$ to write 
\begin{equation}
\langle a_{1,j}^{4}\rangle=3\left(\frac{1}{4\pi}\int_{0}^{\infty}\frac{d\omega}{\omega^2} S(\omega)
\left|y_{1,j}(\omega,\tau)\right|^{2}\right)^2
\end{equation}
which can also be written (to match the other $\xi^4$ terms)
\begin{equation}
\langle a_{1,j}^{4}\rangle=\frac{1}{(4\pi)^2}
\int_{0}^{\infty}\frac{d\omega}{\omega^2}S(\omega)\int_{0}^{\infty}\frac{d\omega'}{\omega'^2}S(\omega')
\left\{3\left|y_{1,j}(\omega,\tau)\right|^{2}\left|y_{1,j}(\omega',\tau)\right|^{2}\right\}
\end{equation}

\newpage
\subsection{$\langle a_{2,i}^{2}\rangle$ - Derived from second-order terms of the Magnus expansion}
\begin{equation}
\langle a_{2,i}^{2}\rangle=\frac{1}{16}\int_{0}^{\tau}dt_{4}\int_{0}^{t_{4}}dt_{3}\int_{0}^{\tau}dt_{2}\int_{0}^{t_{2}}dt_{1}
\langle\eta(t_{1})\eta(t_{2})\eta(t_{3})\eta(t_{4})\rangle s_{2,i}(t_{1},t_{2})s_{2,i}(t_{3},t_{4})
\end{equation}

\subsubsection{Bound}

\begin{align}
s_{2,i}(t_{1},t_{2})s_{2,i}(t_{3},t_{4})=\left(\vec{s}_{1}(t_{2})\times\vec{s}_{1}(t_{1})\right)_{i}
\left(\vec{s}_{1}(t_{4})\times\vec{s}_{1}(t_{3})\right)_{i}
\end{align}
The cross product of two unit vectors is a unit vector, so any component must have a magnitude less than $1$. Thus
\begin{equation}
\langle a_{2,i}^{2}\rangle\leq\frac{3}{4}\xi^4
\end{equation}

\subsubsection{Explicit expression in terms of the noise power spectrum}
\begin{align}
s_{2,i}(t_{1},t_{2})s_{2,i}(t_{3},t_{4})&=\left(\vec{s}_{1}(t_{2})\times\vec{s}_{1}(t_{1})\right)_{i}
\left(\vec{s}_{1}(t_{4})\times\vec{s}_{1}(t_{3})\right)_{i}\\\nonumber
&=\left(\sum_{j,k}\varepsilon_{ijk}s_{1,j}(t_{2})s_{1,k}(t_{1})\right)
\left(\sum_{l,m}\varepsilon_{ilm}s_{1,l}(t_{4})s_{1,m}(t_{3})\right)\\\nonumber
&=\sum_{j,k,l,m}\varepsilon_{ijk}\varepsilon_{ilm}s_{1,j}(t_{2})s_{1,k}(t_{1})s_{1,l}(t_{4})s_{1,m}(t_{3})
\end{align}
\begin{align}
&\langle a_{2,i}^{2}\rangle=\frac{1}{16}\sum_{j,k,l,m}\varepsilon_{ijk}\varepsilon_{ilm}
\int_{0}^{\tau}dt_{4}s_{1,l}(t_{4})\int_{0}^{t_{4}}dt_{3}s_{1,m}(t_{3})\int_{0}^{\tau}dt_{2}s_{1,j}(t_{2})\int_{0}^{t_{2}}dt_{1}s_{1,k}(t_{1})\\\nonumber
&\left\{\langle\eta(t_{1})\eta(t_{2})\rangle\langle\eta(t_{3})\eta(t_{4})\rangle
+\langle\eta(t_{1})\eta(t_{3})\rangle\langle\eta(t_{2})\eta(t_{4})\rangle
+\langle\eta(t_{1})\eta(t_{4})\rangle\langle\eta(t_{3})\eta(t_{2})\rangle\right\}\\\nonumber
\\\nonumber
&\hspace{3cm}=\frac{1}{(4\pi)^2}
\int_{-\infty}^{\infty}\frac{d\omega}{\omega^2}S(\omega)\int_{-\infty}^{\infty}\frac{d\omega'}{\omega'^2}S(\omega')
y_{2,i}(\omega,\omega',\tau)
\end{align}
or
\begin{align}
\langle a_{2,i}^{2}\rangle=\frac{1}{(4\pi)^2}
\int_{0}^{\infty}\frac{d\omega}{\omega^2}S(\omega)\int_{0}^{\infty}\frac{d\omega'}{\omega'^2}S(\omega')
F_{2,a,i}(\omega,\omega',\tau)
\end{align}
where
\begin{equation}
F_{2,a,i}(\omega,\omega',\tau)\equiv
\left\{y_{2,i}(\omega,\omega',\tau)+y_{2,i}(\omega,-\omega',\tau)+y_{2,i}(-\omega,\omega',\tau)
+y_{2,i}(-\omega,-\omega',\tau)\right\}
\end{equation}
and
\begin{align}
\begin{split}
y_{2,i}(&\omega,\omega',\tau)=\frac{\omega^{2}\omega'^{2}}{4}\sum_{j,k,l,m}\varepsilon_{ijk}\varepsilon_{ilm}
\left\{\left(\int_{0}^{\tau}dt_{4}s_{1,l}(t_{4})e^{i\omega' t_{4}}\int_{0}^{t_{4}}dt_{3}s_{1,m}(t_{3})e^{-i\omega't_{3}}\right)\right.\\
&\left.\left(\int_{0}^{\tau}dt_{2}s_{1,j}(t_{2})e^{i\omega t_{2}}\int_{0}^{t_{2}}dt_{1}s_{1,k}(t_{1})e^{-i\omega t_{1}}\right)\right.\\
&\left.+\left(\int_{0}^{\tau}dt_{4}s_{1,l}(t_{4})e^{i\omega' t_{4}}\int_{0}^{t_{4}}dt_{3}s_{1,m}(t_{3})e^{i\omega t_{3}}\right)
\left(\int_{0}^{\tau}dt_{2}s_{1,j}(t_{2})e^{-i\omega' t_{2}}\int_{0}^{t_{2}}dt_{1}s_{1,k}(t_{1})e^{-i\omega t_{1}}\right)\right.\\
&\left.+\left(\int_{0}^{\tau}dt_{4}s_{1,l}(t_{4})e^{i\omega t_{4}}\int_{0}^{t_{4}}dt_{3}s_{1,m}(t_{3})e^{i\omega' t_{3}}\right)
\left(\int_{0}^{\tau}dt_{2}s_{1,j}(t_{2})e^{-i\omega' t_{2}}\int_{0}^{t_{2}}dt_{1}s_{1,k}(t_{1})e^{-i\omega t_{1}}\right)\right\}
\end{split}
\end{align}

\newpage
\subsection{$\langle a_{1,i}a_{3,i}\rangle$ - Derived from third-order terms in the Magnus expansion}
\begin{equation}\label{44}
\langle a_{1,i}a_{3,i}\rangle=\frac{1}{24}\int_{0}^{\tau}dt_{4}\int_{0}^{\tau}dt_{3}\int_{0}^{t_{3}}dt_{2}\int_{0}^{t_{2}}dt_{1}
\langle\eta(t_{1})\eta(t_{2})\eta(t_{3})\eta(t_{4})\rangle s_{1,i}(t_{4})s_{3,i}(t_{1},t_{2},t_{3})
\end{equation}
where
\begin{equation}
\vec{s}_{3}(t_{1},t_{2},t_{3})\equiv\vec{s}_{1}(t_{3})\times\left(\vec{s}_{1}(t_{2})\times\vec{s}_{1}(t_{1})\right)
+\left(\vec{s}_{1}(t_{3})\times\vec{s}_{1}(t_{2})\right)\times\vec{s}_{1}(t_{1})
\end{equation}

\subsubsection{Bound}

\begin{equation}
\left|\langle a_{1,i}a_{3,i}\rangle\right|\leq\frac{\Delta\eta^4\tau^4}{48}=\frac{\xi^4}{4}
\end{equation}

\subsubsection{In Terms of the Noise Power spectrum}
\begin{align}
\langle a_{1,i}a_{3,i}\rangle&=\frac{1}{24}\int_{0}^{\tau}dt_{4}\int_{0}^{\tau}dt_{3}\int_{0}^{t_{3}}dt_{2}\int_{0}^{t_{2}}dt_{1}
s_{1,i}(t_{4})s_{3,i}(t_{1},t_{2},t_{3})\\\nonumber
&\left\{\langle\eta(t_{1})\eta(t_{2})\rangle\langle\eta(t_{3})\eta(t_{4})\rangle
+\langle\eta(t_{1})\eta(t_{3})\rangle\langle\eta(t_{2})\eta(t_{4})\rangle
+\langle\eta(t_{1})\eta(t_{4})\rangle\langle\eta(t_{3})\eta(t_{2})\rangle\right\}\\
&\hspace{3cm}=\frac{1}{(4\pi)^2}
\int_{-\infty}^{\infty}\frac{d\omega}{\omega^2}S(\omega)\int_{-\infty}^{\infty}\frac{d\omega'}{\omega'^2}S(\omega')
y_{3,i}(\omega,\omega',\tau)
\end{align}
Accounting for all terms we generalize
\begin{align}
\langle a_{1,i}a_{3,i}\rangle=\frac{1}{(4\pi)^2}
\int_{0}^{\infty}\frac{d\omega}{\omega^2}S(\omega)\int_{0}^{\infty}\frac{d\omega'}{\omega'^2}S(\omega')
F_{2,b,i}(\omega,\omega',\tau)
\end{align}
where
\begin{equation}
F_{2,b,i}(\omega,\omega',\tau)\equiv
\left\{y_{3,i}(\omega,\omega',\tau)+y_{3,i}(\omega,-\omega',\tau)+y_{3,i}(-\omega,\omega',\tau)
+y_{3,i}(-\omega,-\omega',\tau)\right\}
\end{equation}
and
\begin{align}
\begin{split}
y_{3,i}(\omega,\omega',\tau)=\frac{\omega^2\omega'^2}{4}\left\{
\int_{0}^{\tau}dt_{4}e^{i\omega't_{4}}\int_{0}^{\tau}dt_{3}e^{-i\omega't_{3}}
\int_{0}^{t_{3}}dt_{2}e^{i\omega t_{2}}\int_{0}^{t_{2}}dt_{1}e^{-i\omega t_{1}}s_{1,i}(t_{4})s_{3,i}(t_{1},t_{2},t_{3})\right.\\
\left.+\int_{0}^{\tau}dt_{4}e^{i\omega't_{4}}\int_{0}^{\tau}dt_{3}e^{i\omega t_{3}}
\int_{0}^{t_{3}}dt_{2}e^{-i\omega' t_{2}}\int_{0}^{t_{2}}dt_{1}e^{-i\omega t_{1}}s_{1,i}(t_{4})s_{3,i}(t_{1},t_{2},t_{3})\right.\\
\left.+\int_{0}^{\tau}dt_{4}e^{i\omega t_{4}}\int_{0}^{\tau}dt_{3}e^{i\omega't_{3}}
\int_{0}^{t_{3}}dt_{2}e^{-i\omega' t_{2}}\int_{0}^{t_{2}}dt_{1}e^{-i\omega t_{1}}s_{1,i}(t_{4})s_{3,i}(t_{1},t_{2},t_{3})\right\}
\end{split}
\end{align}

Terms to third order in the control vector  ($s_{3,i}(t_{1}t_{2}t_{3})$) may be evaluated explicitly using the vector identity
\begin{equation}
\vec{a}\times(\vec{b}\times\vec{c})=(\vec{a}\cdot\vec{c})\vec{b}-(\vec{a}\cdot\vec{b})\vec{c}
\end{equation}
to write
\begin{align}
\vec{s}_{3}(t_{1},t_{2},t_{3})=2[\vec{s}_{1}(t_{3})\cdot\vec{s}_{1}(t_{1})]\vec{s}_{1}(t_{2})
-[\vec{s}_{1}(t_{3})\cdot\vec{s}_{1}(t_{2})]\vec{s}_{1}(t_{1})
-[\vec{s}_{1}(t_{1})\cdot\vec{s}_{1}(t_{2})]\vec{s}_{1}(t_{3})
\end{align}
and
\begin{align}
\vec{s}_{3,i}(t_{1},t_{2},t_{3})=2s_{1,i}(t_{2})\sum_{j}s_{1,j}(t_{3})s_{1,j}(t_{1})-
s_{1,i}(t_{1})\sum_{j}s_{1,j}(t_{3})s_{1,j}(t_{2})
-s_{1,i}(t_{3})\sum_{j}s_{1,j}(t_{1})s_{1,j}(t_{2}).
\end{align}
This thus leads to a large proliferation of terms.

\newpage
\section{Summary}
These results may be summarized as follows
\\
\begin{equation}
\langle\mathcal{F}\rangle\simeq1-\sum_{i}\langle a_{1,i}^2\rangle
-\left\{\sum_{i}\left(\langle a_{2,i}^2\rangle+2\langle a_{1,i}a_{3,i}\rangle\right)
-\frac{1}{3}\sum_{i,j}\langle a_{1,i}^{2}a_{1,j}^{2}\rangle\right\}
\end{equation}
\\
In the frequency domain
\begin{align}
\langle\mathcal{F}\rangle=1-\frac{1}{4\pi}\int_{0}^{\infty}\frac{d\omega}{\omega^2} S(\omega)
F_{1}(\omega,\tau)-\frac{1}{(4\pi)^{2}}\int_{0}^{\infty}\frac{d\omega}{\omega^2}S(\omega)\int_{0}^{\infty}\frac{d\omega'}{\omega'^2}S(\omega')
F_{2}(\omega,\omega'\tau)+...
\end{align}
where
\begin{equation}
F_{1}(\omega,\tau)\equiv\sum_{i}\left|y_{1,i}(\omega,\tau)\right|^{2}
\end{equation}
\begin{equation}
F_{2}(\omega,\omega'\tau)\equiv\sum_{i}\left(F_{2,a,i}(\omega,\omega',\tau)+2F_{2,b,i}(\omega,\omega',\tau)\right)
-\frac{1}{3}\sum_{i,j}F_{2,c,i,j}(\omega,\omega',\tau).
\end{equation}

The terms making up $F_{2}(\omega,\omega'\tau)$, and may be written
\begin{equation}
F_{2,a,i}(\omega,\omega',\tau)\equiv
\left\{y_{2,i}(\omega,\omega',\tau)+y_{2,i}(\omega,-\omega',\tau)+y_{2,i}(-\omega,\omega',\tau)
+y_{2,i}(-\omega,-\omega',\tau)\right\}
\end{equation}
\\
\begin{align}
\begin{split}
y_{2,i}(\omega,\omega',\tau)=\frac{\omega^2\omega'^2}{4}\left\{
\int_{0}^{\tau}dt_{4}e^{i\omega't_{4}}\int_{0}^{t_{4}}dt_{3}e^{-i\omega't_{3}}
\int_{0}^{\tau}dt_{2}e^{i\omega t_{2}}\int_{0}^{t_{2}}dt_{1}e^{-i\omega t_{1}}s_{2,i}(t_{1},t_{2})s_{2,i}(t_{3}.t_{4})\right.\\
\left.+\int_{0}^{\tau}dt_{4}e^{i\omega't_{4}}\int_{0}^{t_{4}}dt_{3}e^{i\omega t_{3}}
\int_{0}^{\tau}dt_{2}e^{-i\omega' t_{2}}\int_{0}^{t_{2}}dt_{1}e^{-i\omega t_{1}}s_{2,i}(t_{1},t_{2})s_{2,i}(t_{3}.t_{4})\right.\\
\left.+\int_{0}^{\tau}dt_{4}e^{i\omega t_{4}}\int_{0}^{t_{4}}dt_{3}e^{i\omega't_{3}}
\int_{0}^{\tau}dt_{2}e^{-i\omega' t_{2}}\int_{0}^{t_{2}}dt_{1}e^{-i\omega t_{1}}s_{2,i}(t_{1},t_{2})s_{2,i}(t_{3}.t_{4})\right\}
\end{split}
\end{align}
\\
\begin{equation}
F_{2,b,i}(\omega,\omega',\tau)\equiv
\left\{y_{3,i}(\omega,\omega',\tau)+y_{3,i}(\omega,-\omega',\tau)+y_{3,i}(-\omega,\omega',\tau)
+y_{3,i}(-\omega,-\omega',\tau)\right\}
\end{equation}

\begin{align}
\begin{split}
y_{3,i}(\omega,\omega',\tau)=\frac{\omega^2\omega'^2}{4}\left\{
\int_{0}^{\tau}dt_{4}e^{i\omega't_{4}}\int_{0}^{\tau}dt_{3}e^{-i\omega't_{3}}
\int_{0}^{t_{3}}dt_{2}e^{i\omega t_{2}}\int_{0}^{t_{2}}dt_{1}e^{-i\omega t_{1}}s_{1,i}(t_{4})s_{3,i}(t_{1},t_{2},t_{3})\right.\\
\left.+\int_{0}^{\tau}dt_{4}e^{i\omega't_{4}}\int_{0}^{\tau}dt_{3}e^{i\omega t_{3}}
\int_{0}^{t_{3}}dt_{2}e^{-i\omega' t_{2}}\int_{0}^{t_{2}}dt_{1}e^{-i\omega t_{1}}s_{1,i}(t_{4})s_{3,i}(t_{1},t_{2},t_{3})\right.\\
\left.+\int_{0}^{\tau}dt_{4}e^{i\omega t_{4}}\int_{0}^{\tau}dt_{3}e^{i\omega't_{3}}
\int_{0}^{t_{3}}dt_{2}e^{-i\omega' t_{2}}\int_{0}^{t_{2}}dt_{1}e^{-i\omega t_{1}}s_{1,i}(t_{4})s_{3,i}(t_{1},t_{2},t_{3})\right\}
\end{split}
\end{align}
\\
\begin{align}
F_{2,c,i,j}(\omega,\omega',\tau)
\equiv\left|y_{1,i}(\omega,\tau)\right|^{2}\left|y_{1,j}(\omega',\tau)\right|^{2}
+2Re\left[y_{1,i}(\omega,\tau)y_{1,j}(\omega,\tau)^{*}\right]
Re\left[y_{1,i}(\omega',\tau)y_{1,j}(\omega',\tau)^{*}\right]
\end{align}
with each individual term constituting a single component of $F_{p,2}(\omega, \omega',\tau)$ in the main text.